# Super-Resolution Imaging via Angular Magnification

*Yi Zhou, Dingpeng Liao, Kun Zhang, Zijie Ma, Shikai Wu, Jun Ma, Xuemei Dai, Zhengguo Shang, Zhongquan Wen, Gang Chen,\**

Key Laboratory of Optoelectronic Technology and Systems (Chongqing University), Ministry of Education, School of Optoelectronic Engineering, Chongqing University,174 Shazheng Street, Shapingba, Chongqing 400044, China

Keywords: super-resolution, dielectric metalens, angular magnification

**ABSTRACT:** The far-field resolution of optical imaging systems is restricted by the Abbe diffraction limit, a direct result of the wave nature of light. One successful technological approach to circumventing this limit is to reduce the effective size of a point-spread-function. In the past decades, great endeavors have been made to engineer an effective point-spread-function by exploiting different mechanisms, including optical nonlinearities and structured light illumination. However, these methods are hard to be applied to objects in a far distance. Here, we propose a new way to achieve super-resolution in a far field by utilizing angular magnification. We present the first proof-of-concept demonstration of such an idea and demonstrate a new class of lenses with angular magnification for far-field super-resolution imaging. Both theoretical and experimental results demonstrate a more than two-fold enhancement beyond the angular-resolution limit in the far-field imaging. The proposed approach can be applied to super-resolution imaging of objects in far distance. It has promising potential applications in super-resolution telescopes and remote sensing.

# ■ INTRODUCTION

Due to the spreading nature of light wave propagation, the optical image of an ideal infinitely small point obtained through optical devices or systems becomes a central, bright disk of infinite size surrounded by concentric rings [1]. This sets a fundamental restriction on the resolution of optical imaging, which was first proposed by Ernst Abbe [2]. According to Rayleigh criterion, the diffraction limit restricts the ability of optical systems to resolve objectives smaller than $0.61\lambda/NA$, where $\lambda$ and NA are wavelength and numerical aperture. To break through this limitation, great efforts have been made in the past half century, among which near-field probing [3] was firstly proposed. By utilizing non-propagating high-frequency components in the optical field near the objective surface, small details beyond the diffraction limit have been observed through a scanning probe moving above the surface at a working distance much less than a wavelength. Hyperlenses and metalenses can convert near-field waves into propagation waves in specially designed high-effective-index metamaterials for far-field imaging, however it working distance is in sub-wavelength scale and cannot be applied to imaging for objects in far distance. [4,5] Far-field super-resolution optical microscopy has also experienced a rapid development, which can be divided into three major categories, i.e., single molecule localization, point-spread-function engineering and frequency shifting. Single molecule localization microscopy can achieve a resolution of 10 nm [6] by using photoactivation or photoswitching of single fluorophores and position determination; stimulated emission depletion microscopy can compress the effective point-spread-function far beyond the diffraction limit and achieve a resolution of 2.4 nm [7] by utilizing non-linear effect in the competition between stimulated emission and autofluorescence; structured light illumination microscopy can realize a resolution of λ/5 [8] by shifting the higher frequency information into propagation wave which is used to reconstruct the high resolution image of an objective. However, the first two approaches largely rely on fluorescence labeling, and the third one depends on reconstruction algorithms. Physically, these approaches essentially work on compressing the effective point-spread-function of the whole system, which involves not only the optical system itself but also the objective under test. Microspheres-assisted super-resolution microscopy can realize far-field imaging by utilizing increased effective numerical aperture, however its working distance is limited to several micrometers in optical domain and cannot be applied to imaging for objects in far distance too. [9,10] The recent proposal of the concept of optical superoscillation [11-13] provides an alternative way to achieve far-field super-resolution by engineering the point-spread-function of optical devices or systems. According to the theory, an arbitrary small point-spread-function can be realized by carefully designing the complex optical transmittance function. Various types of superoscillation optical devices [14-19] have been demonstrated, showing promising potentials in realizing super-resolution telescopes [20-22] and far-field label-free super-resolution microscopy. [23-26] However, the inevitable strong sidebands significantly limit the field of view in the realistic applications. Especially in telescopes, the strong sidebands result in a comparatively weak super-resolution image surrounded by strong sidebands.

In traditional optics, the angular magnification of optical lens is equal to 1, when they work in a single medium, for example in air. Due to this fact, as having been done in the past decades, the natural and direct way to overcome the diffraction limit is to compress the effective size of the point-spread-function. In contrast, another possible solution is to enhance the angular magnification without increasing the effective size of the point-spread-function. Here, we have proposed such an approach to realizing far-field super-resolution imaging by angular magnification. The validity of the concept has been theoretically and experimentally proved and demonstrated with properly designed lenses with angular magnification greater than 1. We also prove that the strong sideband can be eliminated in the super-resolution image by using the proposed approach, showing its obvious advantages over super-resolution point-spread-function engineering. The proposed approach can be applied to super-resolution imaging of objects in far distance. It has promising potential applications in super-resolution telescopes and remote sensing.

## ■ THEORY ON SUPER-RESOLUTION IMAGING VIA ANGULAR MAGNIFICATION

**Figure 1a** presents the schematic diagram of optical imaging of two ideal points $A$ and $B$ through a conventional optical lens. In the object space, the spacing between the two points $A$ and $B$ is $d_o$; the angle between the chief rays of the two points is $\theta$; the object distance is $l_0$; the refractive index is $n$. In the image space, the spacing between the two image $A'$ and $B'$ is $d_i$; the angle between the chief rays of the two image points is $\theta'$; the image distance is $l_i$; the size of the point-spread-function is $d$; the refractive index is $n'$. The lens radius is $R$. Generally, when the refractive indices are equal, i.e. $n=n'$, the angles in both sides of the lens are also equal, i.e. $\theta=\theta'$.

According to the Rayleigh diffraction limit, the achievable minimum size of the point-spread-function on the image plane is given by $d=0.61\lambda/NA$, where $NA=\sin(\operatorname{atan}(R/l_i))$, and the two points $A$ and $B$ cannot be distinguished when the spacing between the two image spots is less than $d$, as shown in **Figure 1a**. Therefore, the corresponding angular resolution in the object space is restricted to $\delta\theta_{DL} = 2\mathrm{atan}(0.5d/l_i)$, which can be rewritten as equation (1). If $l_i$ is much greater than $R$, it can be further simplified as equation (2).

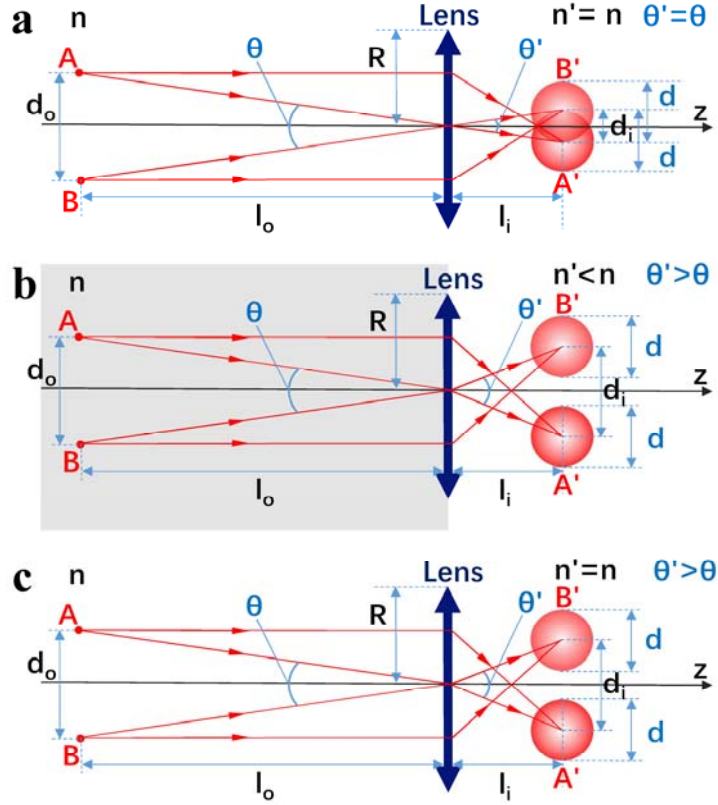

**Figure 1.** (a) Imaging using the conventional optical lens with angular magnification $K=1$; (b) imaging using the conventional medium-immersion optical lens with angular magnification $K>1$; (c) imaging using the proposed super-resolution lens with angular magnification $K>1$.

$$\delta\theta_{DL}(NA) = 2\mathrm{atan}(0.5 \times 0.61\lambda/(NA \cdot l_i)) \qquad (1)$$

$$\delta\theta_{DL}(0) = 1.22\lambda/D \qquad (2)$$

where $D$ is the diameter of the lens, i.e. $D=2R$. In a more general case, if the two angles are not equal and $\theta'$ is greater than $\theta$, or $\theta'=K\theta$ ($K > 1$), then the angular resolution $\delta\theta_K$ in the object space should be modified by adding an additional factor $1/K$ to equation (1), as expressed in equation (3). Similarly, it can be simplified as equation (4) when $l_i$ is much greater than $R$.

$$\delta\theta_K(NA) = 2\mathrm{atan}(0.5 \times 0.61\lambda/(K \cdot NA \cdot l_i)) \qquad (3)$$

$$\delta\theta_K(0)=1.22\lambda/(K\cdot D) \qquad (4)$$

Therefore, the corresponding resolution can be enhanced by a factor of *K*. Such angular magnification can be realized by making the refractive index in the object space higher than that in the image space, [27] as shown in **Figure 1b**. This technique has been implemented in medium-immersion-microscopy, where the object space is immersed in a transparent medium with a high refractive index greater than the medium of imaging space. However, this enhancement of the angular resolution demands a higher refractive index in the object space, which thus poses severe limitations to the use of medium-immersion imaging in many practical applications when the objects are in far distance.

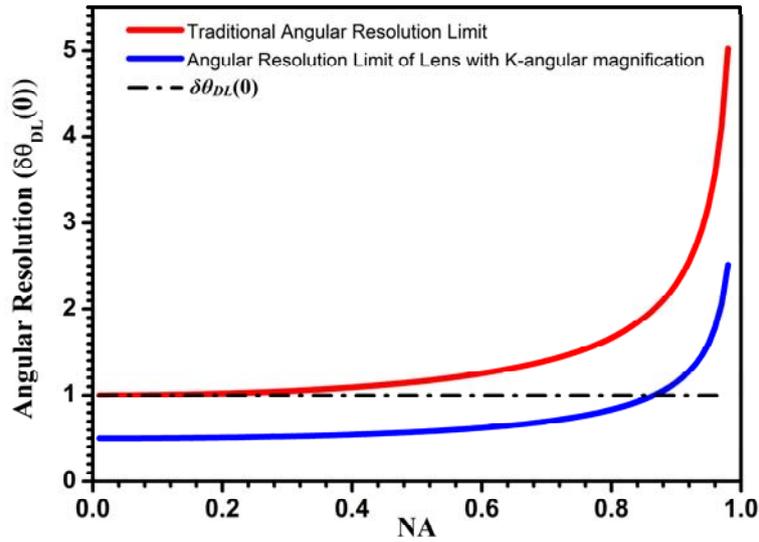

**Figure 2**. Comparison of the different angular-resolution-limits. The traditional angular-resolution-limit at different numerical aperture is given by the red curve; the angular-resolution-limit of a lens with angular magnification of *K*=2 is given by the blue curve; the limitation of the traditional angular-resolution-limit $\delta\theta_{DL}(0)$ is presented as the numerical aperture approaching zero, given by the black dash-dotted line.

**Figure 2** gives the comparison of the different angular-resolution-limits. In the figure, the traditional angular-resolution-limit $\delta\theta_{DL}(NA)$ at different numerical apertures is given by the red curve; the angular-resolution-limit $\delta\theta_K(NA)$ of a lens with angular magnification of *K*=2 is given by the blue curve; the limitation of the traditional angular-resolution-limit $\delta\theta_{DL}(0)$ is presented as the numerical aperture approaching zero, given by the black dash-dotted line. It is found that the angular-resolutions given by the two curves increase as the numerical aperture decrease, and both achieve their minimum angle limitations of $\delta\theta_{DL}(0)$ and $\delta\theta_K(0)$ respectively at NA=0.

Is it possible to achieve angular magnification without using medium-immersion approaches? As shown in **Figure 1c**, does such a type of lenses which has the ability of angular magnification $K > 1$ exist even when the refractive indices are equal in both object and image spaces? Here we report the first experimental and numerical demonstration of the possibility to realize such lenses for super-resolution imaging by angular magnification. Experimentally, we can resolve the object details at $1/K$ of the diffraction limit imposed by the conventional optical systems.

## ■ SUPER-RESOLUTION LENSES OF ANGULAR MAGNIFICATION

**SLAM Singlet with Large Numerical Aperture.** To prove our concept, a SLAM singlet with $K = 1.9487$ has been designed by using angular spectrum method [28] and optimization algorithm [29]. As shown in **Figure 3a**, to realize the lens phase profile, an a-Si Pancharatnam–Berry phase meta-atom [30] with a unit size of $P \times P$ has been adopted for the wavelength of 632.8 nm. It is an a-Si cubic block with a height of $H$, length of $L_s$ and width of $L_f$ on the top of a $SiO_2$ substrate. For a circularly polarized incident wave, it introduces a phase shift of $\varphi=2\alpha$ in the cross-polarized emergent wave with a rotation angle of $\alpha$. Based on the a-Si meta-atom, a phase-only transmission function has been optimized for a lens with given focal length of $f = 60\lambda$ and radius of $R = 240\lambda$ at wavelength of $\lambda$=632.8 nm. The lens numerical aperture is NA=0.97. The corresponding angular resolution limits are $\delta\theta_{DL}(NA) = 0.6005°$ at NA = 0.97 and $\delta\theta_{DL}(0) = 0.1456°$ at NA=0, respectively.

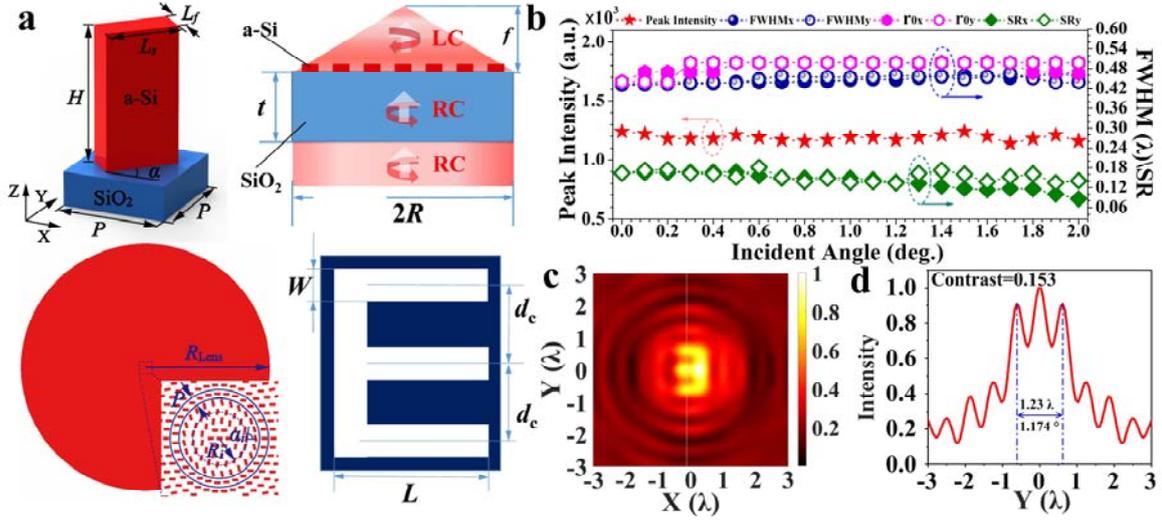

**Figure 3.** Simulation results of the SLAM singlet. (a) the diagram of the a-Si meta-atom, the diagram of the side view and top view of the SLAM singlet, and the letter "E" used as the object for imaging; (b) the peak intensity (red stars), FWHW (royal spheres), first-zero $r_0$ (Magenta hexagons) and sidelobe ratio (Olive diamonds) of the focal spot with different incident angles, where solid and open markers denote the results in x-direction and y-direction respectively; (c) the incoherent image of letter "E" at the object distance of $l_o$=18 mm and image distance of $l_i$=37.97 μm under the incoherent illumination at wavelength of 632.8 nm; (d) the intensity distribution along the white-dashed line in (d).

As shown in **Figure 3a,** the proposed SLAM singlet consists of a series of ring belts with width of $P$. The central radius of the $i$-th ring belt is denoted by $R_i$, which is equal to $i \times P$. The meta-atoms within the same ring belt have the same phase value. For right-circular (RC) polarization incident wave, the lens converts the polarization into left-circular (LC) polarization and focuses the emergence wave on the focal plane. In the design, the angular spectrum method [28] is used to calculate the diffraction pattern of the lens, and the particle-swarm optimization algorithm [29] is used to find the desired phase profile of the lens. The phase profile is optimized in a way that guarantees the focal spot is (FWHM, full-width-at-half-maximum) no greater than $FWHM_{max}=0.5\lambda$ and sidelobe ratio (SR, the ratio of the maximum sidelobe intensity to the central lobe intensity) is no greater than $SR_{max} = 0.2$ on the focal plane under the illumination of the plane waves at incident angles within the given field-of-view (FOV) of 4° in the object space which correspond to a FOV of 7.7948° in the image space for angular magnification $K = 1.9487$.

The critical constraint is also applied to achieve the linear-relationship between the incident angle $\theta_i$ and the emergence angle $\theta_e$, i.e. $\theta_e = K\theta_i$. The optimized phase profile $\varphi(r)$ is presented in **Table 1**. **Figure 3a** also illustrates a letter of "E" used as the object for imaging. The linewidth of the letter is $W$, the center-to-center distance between two neighboring lines is $d_c$, and the width of the entire letter is $L$.

Table 1. The phase distribution of the SLAM singlet along the radial direction, where the phase number is given by Ni in the base-32 numerical system for the meta-atom at the location $(i-1)P$ away from the lens center; the corresponding phase value of $\varphi_i$ is equal to $2\pi N_i/32$.

| No. of Ring Belt | $N_i$ |
|---|---|
| #1~#169 | 1802G2801511T100TMS1D00G2650M0023006E00100000943A00O00N2000041000A000304000706U00905C0100022809B000200A9F6S064E60020001000B10D0500M00040000K20G30017020E21000202006 16020I |
| #169~#338 | 0F00B00D0K50110700205NF3030I10C0040B0F8R70F0080B1B703000N00F409071140705075060B6E5440227001O4000CAF23C215B6B8C42266020AU2C0DREA1506260C154003K0C000207050000 90I0000700870 |
| #339~#506 | 6250C0007000000200000010000000000070GO420N0270KC0017N00E01U070710010000025R000F009D000000030020000 00B0D070Q20JC0F8K004F00E0000N0000H00900R4V73E0DVH1N20702033M0F80000B00N |

Numerical simulations have been conducted to obtain the intensity distribution on the focal plane for 21 different incident angles with an equal interval of 0.1° between 0 and 2°. It is found that there is only one single spot on the designed focal plane at each simulated angle and the displacement of the focal spot monotonically increases with the increase of the incident angle. **Figure 3b** depicts focal spot intensity, FWHM, first-zero of the intensity pattern $r_0$, SR and $\theta_e$ with respect to the incident angle $\theta_i$. It is seen that both FWHM and $r_0$ are smaller than the Rayleigh's diffraction limit of $0.61\lambda/NA = 0.629\lambda$ and the SR is smaller than 0.181, for all incident angles. It is noted that the focal spot intensity only fluctuates slightly in the range of 8% of the maximum value within the designed field-of-view. According to the result, the emergence angle $\theta_e$ varies linearly with the incident angle $\theta_i$ and the angular magnification is $K=1.9487$.

**Figure 3c** presents the incoherence imaging result of letter "E" at illumination wavelength of 632.8 nm, where the object distance is $l_o$ = 18 mm and the image distance $l_i$ =37.97 μm. The width of the entire letter is $L$ = 120 μm, the linewidth of the letter is $W$ = 24 μm and the center-to-center distance between two neighboring lines is $d_c$ = 80 μm which corresponds to an angle of $\delta\theta$ = 0.255º at the object distance of $l_o$ = 18 mm. **Figure 3d** plots the intensity distribution along the white-dashed line in **Figure 3c.** Because the Rayleigh resolution criterion requires a minimum contrast of 10.5% [28], the image of the letter of "E" can be clearly resolved due to the image contrast of 15.3%. Therefore, the SLAM singlet can achieve an angular resolution of 0.255º, which is only 0.425 time of the angular resolution limit $\delta\theta_{DL}(NA)$ = 0.6005º at NA = 0.97. The angular resolution shows more than two-fold enhancement, resulting from both sub-diffraction spot size and angular magnification. The spot size is approximately 0.787 times of the Rayleigh's diffraction limit, while the angular magnification $K$=1.9487. Therefore, ideally, the angular resolution can achieve approximately 0.404 (0.787/K), which is slightly smaller than the simulation result. This difference is due to the fact that the SR of the focal spot of the SLAM singlet is greater than that of Airy spot. It is clearly seen that the angular magnification plays a major role in the enhancement of the angular resolution, which proves the super-resolution imaging ability of the proposed SLAM singlet.

**SLAM Doublet with Small Numerical Aperture.** Although the SLAM singlet shows an enhancement of angular resolution by a factor of 2.353 compared with that of a perfect conventional imaging lens with the same numerical aperture of NA=0.97, its resolution remains above the theoretical limitation with the same radius and NA close to zero, as illustrated in **Figure 2**.

To further verify our concept at smaller NA, a SLAM doublet has been designed with an angular magnification of $K=1.88345$ and a numerical aperture of NA=0.44 at wavelength of $\lambda=632.8$ nm using the same method. As presented in **Figure 4 a**, the SLAM doublet consists of two Pancharatnam–Berry metasurfaces with the same radius of $R$ on both sides of the glass substrate with a thickness of $t$. The refractive index of the glass substrate is 1.457 at wavelength of 632.8 nm. For the focal length of $f=816\lambda$ and radius of $R=400\lambda$, the optimized substrate thickness of $t$ is 675 μm. The corresponding angular resolution limit is $\delta\theta_{DL}(NA) = 0.0973°$ at NA = 0.44 and $\delta\theta_{DL}(0) = 0.0874°$ at NA=0, respectively. **Table 2** displays the optimized phase distributions of the two functional metasurfaces in 32-based digitals.

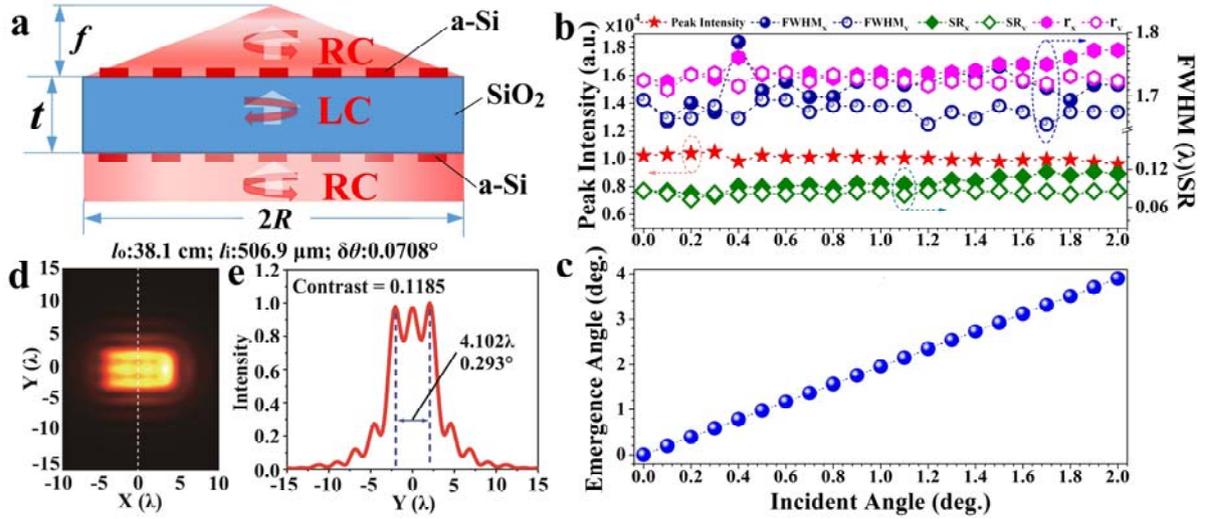

**Figure 4.** Simulation results of the SLAM doublet. (a) the diagram of the proposed lens; (b) the peak intensity $I_{peak}$ (red stars), FWHW (royal spheres), $r$ (Magenta hexagons, the half-width at intensity of $0.01I_{peak}$) and sidelobe ratio (Olive diamonds) of the focal spot at different incident angles, where solid and open markers denote the results in x-direction and y-direction respectively; (c) the relation between the emergence angle $\theta_e$ and the incident angle $\theta_i$; (d) the image of letter "E" at the object distance of $l_o$=38.1 cm and image distance of $l_i$=506.9 μm under the incoherent illumination at wavelength of 632.8 nm; (e) the intensity distribution along the white-dashed line in (d).

**Figure 4b** shows the major focusing parameters, i.e. peak intensity $I_{peak}$, FWHM, $r$ (the half-width at intensity of $0.01I_{peak}$) and SR, on the designed focal plane at different incident angles. It is seen that the both FWHM and $r$ are smaller than $1.8\lambda$, which is slightly greater than the Rayleigh diffraction limit $0.61\lambda/NA = 1.386\lambda$. **Figure 4c** plots the relation between the emergence angle $\theta_e$ and the incident angle $\theta_i$. The emergence angle clearly shows an excellent linear relationship with the incident angle and the angular magnification $K$ is approximately 1.88345. **Figure 4e** presents the incoherence imaging result of letter "E" at illumination wavelengths of 632.8 nm. The object distance is $l_o = 38.1$ cm and the image distance is $l_i = 506.9$ μm. The width of the entire letter is $L = 1884$ μm, the linewidth of the letter is $W = 80$ μm and the center-to-center distance between two neighboring lines is $d_c = 471$ μm, which corresponds to an angle of $\delta\theta = 0.0708°$ at the given object distance of $l_o = 38.1$ cm. This angle is only 0.707 times of angular resolution limit of $0.09734°$ at NA=0.44 ($0.61\lambda/(NAf)$), and it is smaller than the angular resolution limit of $0.08737°$ at NA=0 ($1.22\lambda/D$). **Figure 4d** plots the intensity distribution along the white-dashed line in Figure 3d, and it shows that the image of the letter of "E" can be clearly resolved with a contrast of 11.85%, which is greater than the contrast value required by the Rayleigh resolution criterion. Therefore, the SLAM doublet proves the super-resolution ability of the lens with angular magnification at small numerical aperture.

**Table 2. The phase distribution of the SLAM doublets along the radial direction, where the phase number is given by Ni in the base-32 numerical system for the meta-atom at the location $(i-1)P$ away from the lens center; the corresponding phase value of $\varphi_i$ is equal to $2\pi N_i/32$.**

| No. of Ring Belt | $N_i$ (front lens) | $N_i$ (rear lens) |
|---|---|---|
| #1~#281 | 0008V02220T000J0000000010VJ00000U00J000000000S000000000000000N0QS00000B0OJ00Q000J0040F7FF0LB0H0FDB000H022002100GG30000G400I40000J000D0HN020V0KT00JT07GA0053U000A0000000NH0G91T809D0080B00I0T500PLA022802K1T0420SA0000380SQ20F1MDFR96I4003SA00N0080J00M08V7LVQ7P00CM40K2002CGP0009H7S03I42 | 0L0H0000000000000000000000I5000000000000001000504000000000010000000 0T0000000000000000M000000000000000 000000000000C000000001100000000000 00L00Q00S0000000020000000000000000 000000000000000000000000000000000000 000000000B00200000000000600000200 0000000000000000000010000000000000 00000000000 |
| #169~#338 | 240V702FDVH7000C000C0AGBOH3UG2 0O07300EV4J8DF81J3B0G0V200NC0E0T 0U23Q12000020000000092000R0TD000U 0RDC0VIO0E1000020JH0940370110I0003 UP00I30224ADN05J00LA00E010A00000F 400K0HKF48KQGGR10010F0050T000N0 F2V0J49200M000IQ0S20C27100ON0F0FD D0051090R1OH001200I0F0B020L20ICJN0 N00IOB87DI0001KFDFA7V | 00001000000T0000000000000000000000 010000V0000000000000000000000000000 0000000000000000000000000000000002 0000000200000000000000000000000000 0000000100000000000000000000000000 000000000000000000000000000000000R0 00000000000000000000000D000000Q00 02000000000M00S000B00800F000000000 0770U702010 |
| #339~#506 | VJF010OD00O0B080DCK300I0N0600L89 PJTC12UQGVAB38S1100F9I03GG6N0NQ C07A180000NJQLAG2102N004F66V3020 0FJEA72N00NFBG93804NC0J8G25V013 MBIAB60N0NJ0HEA800NQKIAGM243N 1LH2386000LJFC9A50FK0FGA10V00EF7 2S40N00F01D6100JOI026204JJ7D921N00J K1AV000NK2C88000MF0B960U000DB12 0000F775V00FHEA6100NFED71V0P0 | 00LA0001AM0001E00F0Q16000000400 24032O41007000C000000EF000000030J 0G0070000E0002000000000001F0D00B 00000E010010F2G0DB00000000000E00 020F00D0GF0000G020006020000000R0 0N00002A04006004000080100E70AD77 0DA0070F0ED009047S00000070RBF0A 0E80B0800I00PN1RN0DK0QN00L0070 02RJ15800RS00M0TM01200DV0N000 |

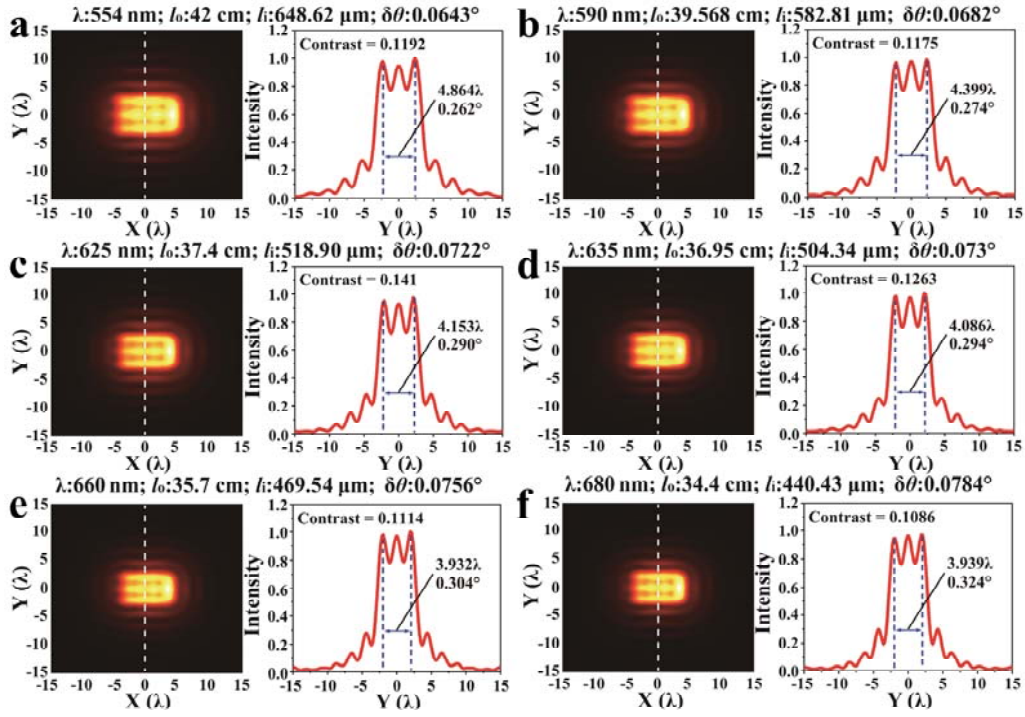

**Figure 5.** Simulation results of broadband incoherent imaging with the SLAM doublet. The image of letter "E" at different wavelengths of (a) 554 nm, (b) 590 nm, (c) 625 nm, (d) 635 nm, (e) 666 nm and (f) 680 nm, where the left is the two-dimensional intensity distribution on the corresponding image plane and the right is the intensity distribution along the corresponding white-dashed line. The object distance and image distance are denoted by $l_o$ and $l_i$ respectively.

For the SLAM doublet, simulations have also been conducted to investigate its broadband super-resolution imaging performance at different wavelengths of 554 nm, 590 nm, 625 nm, 635 nm, 666 nm and 680 nm. The simulation results are presented in **Figure 5**. The corresponding object distance and image distance ($l_o$, $l_i$) are (42 cm, 1170.8λ=648.62 μm), (39.568cm, 987.8λ=582.81 μm), (37.4cm, 830.2λ=518.90 μm), (36.95 cm, 794.2λ=504.34 μm), (35.7 cm, 705.0λ=469.54 μm), and (34.4 cm, 647.69λ=440.43 μm) respectively. The angular resolutions can achieve 0.0643º, 0.0682º, 0.0722º, 0.073º, 0.0756º and 0.0784º respectively, which are smaller than the corresponding angular resolution limit $\delta\theta_{DL}$(NA) of 0.0821, 0.0888, 0.0960, 0.0981, 0.1045 and 0.1083. They are even smaller than the angular resolution limit $\delta\theta_{DL}$(0) of 0.07649, 0.08146, 0.08629, 0.09227, 0.09195, 0.09389º respectively at NA=0 (1.22λ/D). Therefore, the SLAM doublets can achieve super-resolution imaging in a broadband wavelength range of 126 nm.

## ■ EXPERIMENTAL RESULTS AND DISCUSSIONS

For easy fabrication, a SLAM singlet based on the above design has been fabricated with conventional nanofabrication techniques. The a-Si meta-atom has been optimized at wavelength of 632.8 nm. The pitch of the meta-atom is $P = 300$ nm, the length, width and height of the meta-atom are $L = 200$ nm, $W = 116$ nm and $H = 330$ nm, respectively. The phase $\varphi(r_{ij})$ at point of $r_{ij}$ can be realized through rotating the a-Si block by an angle $\alpha=\varphi(r_{ij})/2$. To experimentally prove the validity of our idea, super-resolution imaging experiments have been conducted for incoherent illumination.

**Experimental Setup.** As shown in **Figure 6**, a LED is used as the incoherent illumination source for imaging. A pair of linear polarizer (LP) and quarter-wave plate (QWP) is used to generate desired circular polarization. By utilizing a positive lens, the light is then collected and focused on to a glass plate with an array of letter "E", which is used as the object for imaging. The fabricated meta-lens is mounted on a three-dimensional linear translation stage (LNR25, Thorlabs Inc.) right after the object. The distance between the object and the SLAM singlet is kept greater than the designed focal length. The image formed on the image plane of the SLAM singlet is acquired with a high-numerical and high-magnification microscope, which is composed of an infinite objective lens (CF Plan 150×/0.95, Nikon), a one-dimension nanopositioner (P-721.CDQ, Physik Instrumente), a tube lens (ITL200, Thorlabs, Inc.) and a high-resolution digital camera (acA3800-14μm, Basler, Inc.). Between the objective lens and the tube lens, another pair of QWP and LP is used to filter out the unwanted polarization. The objective lens is mounted on the nano-positioner, which is used to conduct the scan in the z-axis to obtain the two-dimensional optical intensity distribution at different z coordinates. The image of the diffraction pattern is recorded through the digital camera. The "E" letter on the object has the entire width of $L = 120$ μm, linewidth of $W = 24$ μm and the center-to-center distance between two neighboring lines $d_c = 80$ μm.

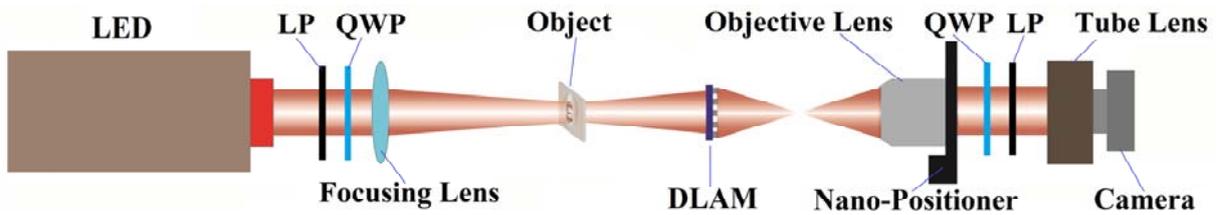

**Figure 6.** The setup of the imaging experiments, where LPs and QWPs are linear polarizers and quarter-wave plates, respectively.

**Results and Discussions.** In the experiments, for the object distances of $l_o = 12$ mm, 13 mm, 14 mm and 15 mm, the images were obtained through SLAM singlet at wavelength of 625 nm. These images are depicted in **Figure 7a**, in which the corresponding angles between the two neighboring horizontal lines in character "E" are 0.382º, 0.353º, 0.327º and 0.306º, respectively.

The corresponding image distances are $l_i$ = 40.0 μm, 37.5 μm, 35.7 μm and 39.1 μm respectively. According to our theory, as given by equation 3, the minimum angle that can be resolved by the SLAM singlet is 0.308º, which is roughly half of the angular resolution limit of 0.613º allowed by a conventional optical lens with the same focal length of 60λ and radius of 240λ. According to the results presented in **Figure 7a**, the character "E" can be clearly identified in all four cases. **Figure 7b** plots the intensity distributions along the dashed lines in **Figure 7a**, based on which the contrast has been calculated for each case. The corresponding contrast values are 0.22, 0.20, 0.168 and 0.158, respectively. All the contrast values are greater than the minima contrast value of 0.105 for the resolvable condition required by Rayleigh criterion. Therefore, the experiments give the best resolution of 0.306º, which is slightly smaller than the predicted 0.308º by equation 3. This is because a conventional Airy spot has a spot size of 0.61λ/NA and sidelobe ratio less than 1.57%, while the focal spot of the proposed metalens has a smaller spot size (FWHM) of approximately 0.47λ and a larger sidleobe ratio of approximately 20% in the experiments. It is found that the peak-to-peak distances between two outer lines of the image of letter 'E' are 1.501λ, 1.437λ, 1.334λ and 1.366λ for the above four different cases, which correspond to angles of 1.344º, 1.372º, 1.291º and 1.254º in the image space. The corresponding angular magnifications are 1.759, 1.943, 1.974 and 2.109 respectively, which are close to the theoretical design value of 1.9487. **Figures 7c and 7d** give the corresponding numerical simulation results of a single character "E" with the same geometrical size and under the same object distances. The corresponding image distance is 39.1 μm for the largest contrast value. The simulation results give better contrast values of 0.22, 0.361, 0.407 and 0.408 for each case. Compared with the numerical results, the lower contrast values obtained in the experiments are due to two reasons. Firstly, it is caused by the experimental and fabrication errors; Secondly, there are influences caused by neighboring characters in the array of letter "E" in the experiments, while there is only single letter "E" in the numerical simulations.

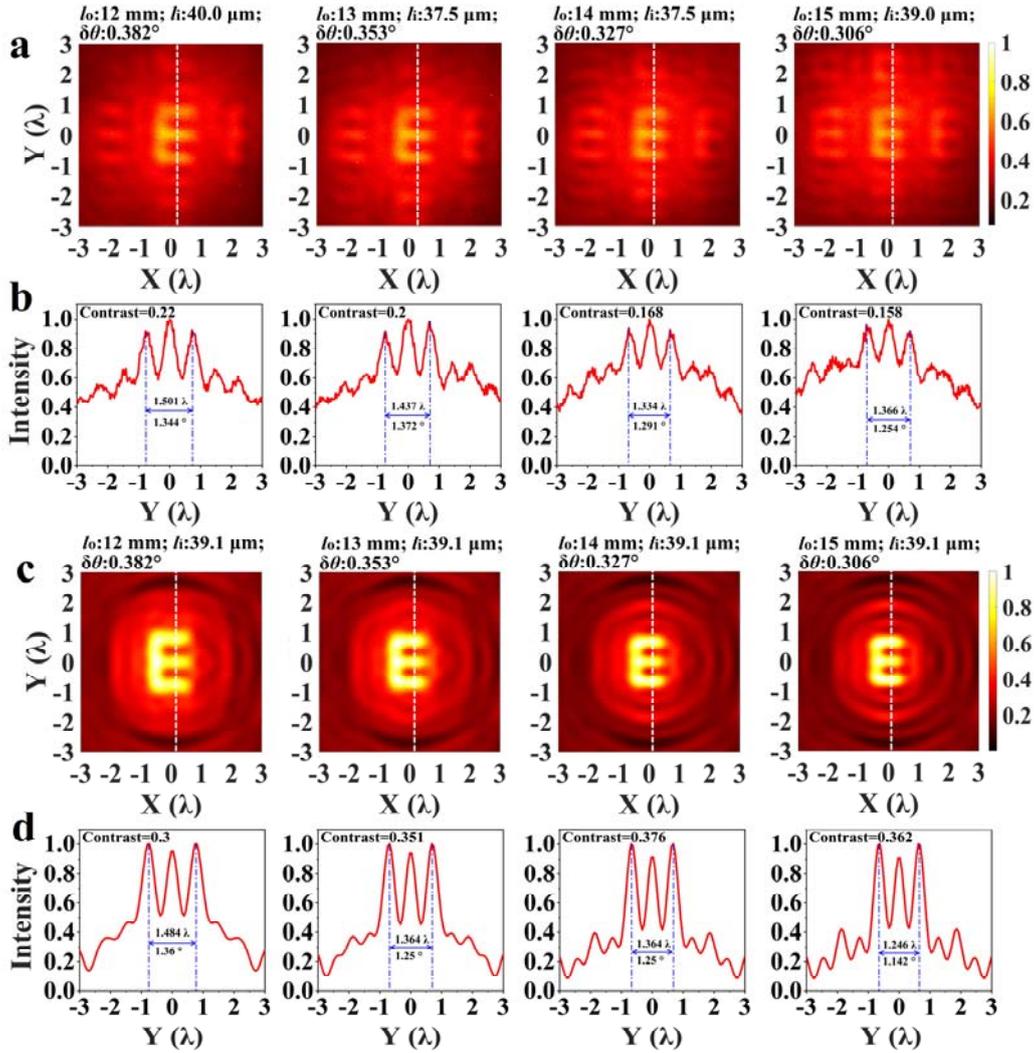

**Figure 7.** The imaging results of the proposed SLAM singlet at different objective distances of $l_o$. (a) the experimental imaging results; (b) the intensity curve along the dashed line in (a); (c) the simulation imaging results; (d) the intensity curve along the dashed line in (c).

As indicated in **Figures 7c and 7d**, the peak-to-peak distances between two outer lines of the image of letter 'E' are $1.484\lambda$, $1.364\lambda$, $1.364\lambda$ and $1.264\lambda$ for the above four different cases, which correspond to angles of 1.36º, 1.25º, 1.25º and 1.142º in the image space respectively. The corresponding angular magnifications are 1.78, 1.77, 1.91 and 1.87 respectively, which are close to the theoretical design value of 1.9487. It is also found that the size of the image decreases as the objective distance increases in both the experimental and numerical results, which is similar to that observed in conventional imaging lens. Interestingly, although the angular resolution achieves more than two-fold enhancement, unlike the super-resolution telescope based on the point-spread-function engineering and the concept of superoscillation, where a weak image is inevitable surrounded by a bright huge sideband, [20-23] there is no obvious sideband appearing in the image plane in present case. Therefore, both the experimental and numerical results prove the validity and advantages of the proposed super-resolution imaging by angular magnification.

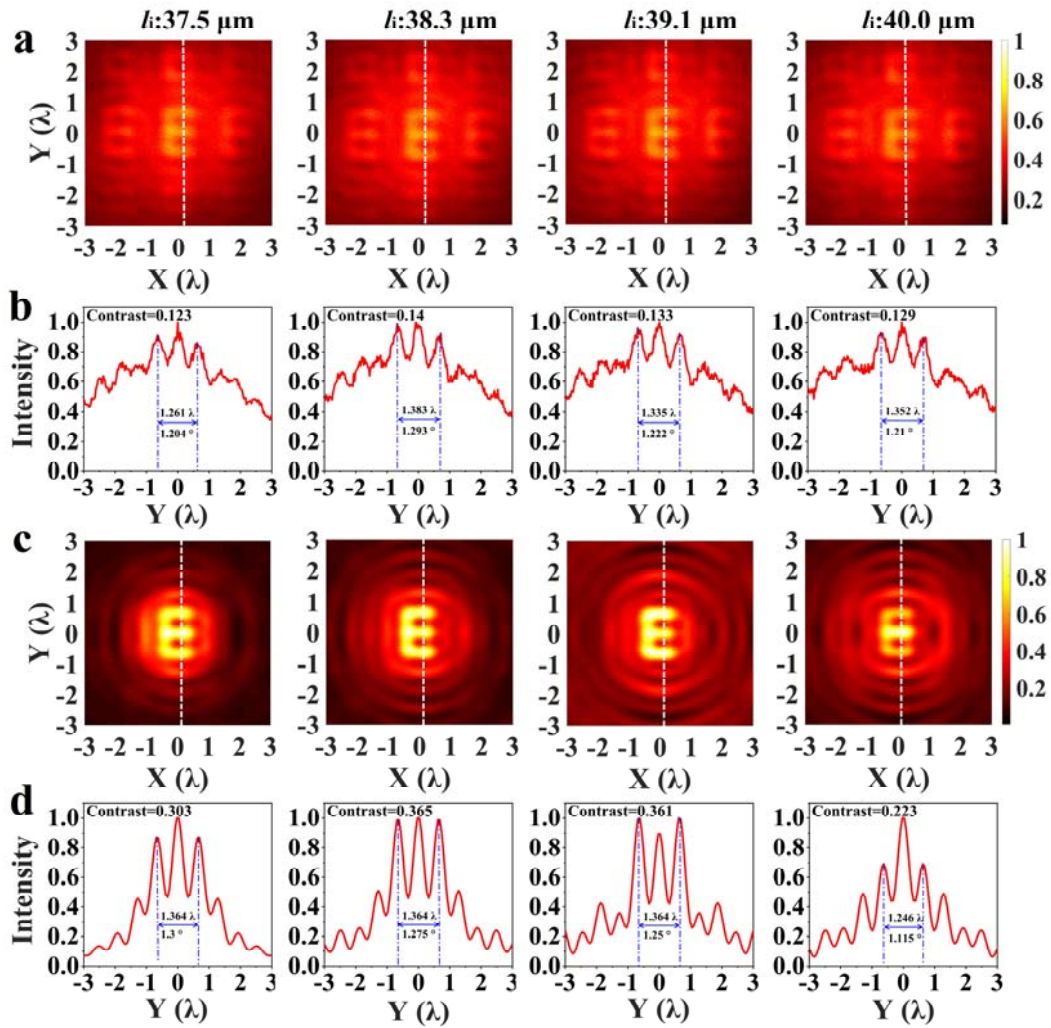

**Figure 8.** The imaging results of the proposed SLAM singlet at different image distances of li. (a) the experimental imaging results; (b) the intensity curve along the dashed line in (a); (c) the simulation imaging results; (d) the intensity curve along the dashed line in (c).

To investigate the tolerance of the image position, the on-axis scanning of the objective lens was conducted. The images have been acquired at different image distances of $l_i$ = 37.5 μm, 38.3μm, 39.1 μm and 40.0 μm for a fixed object distance of $l_o$= 1500μm, corresponding to the actual minimum resolvable angle of 0.306º of the SLAM in the experiment. The experimentally obtained images are presented in **Figure 8a**. In **Figure 8b**, the intensity distributions along the dashed lines in **Figure 8a** are plotted for those images. It is found that, in the range of the object distance between 37.5 μm and 40 μm, the images obtained with the lens can be clearly resolved with contrast values of 0.123, 0.14, 0.133 and 0.129 respectively, which are greater than the Rayleigh criterion contrast value of 0.105. It indicates a depth-of-focus of 2.5 μm (4λ), which is unexpectedly large for a lens with NA of 0.97. **Figure 8c and 8d** are the corresponding simulation results for comparison. Consequently, for the two reasons discussed above, the images obtained by the numerical simulations show better contrast values.

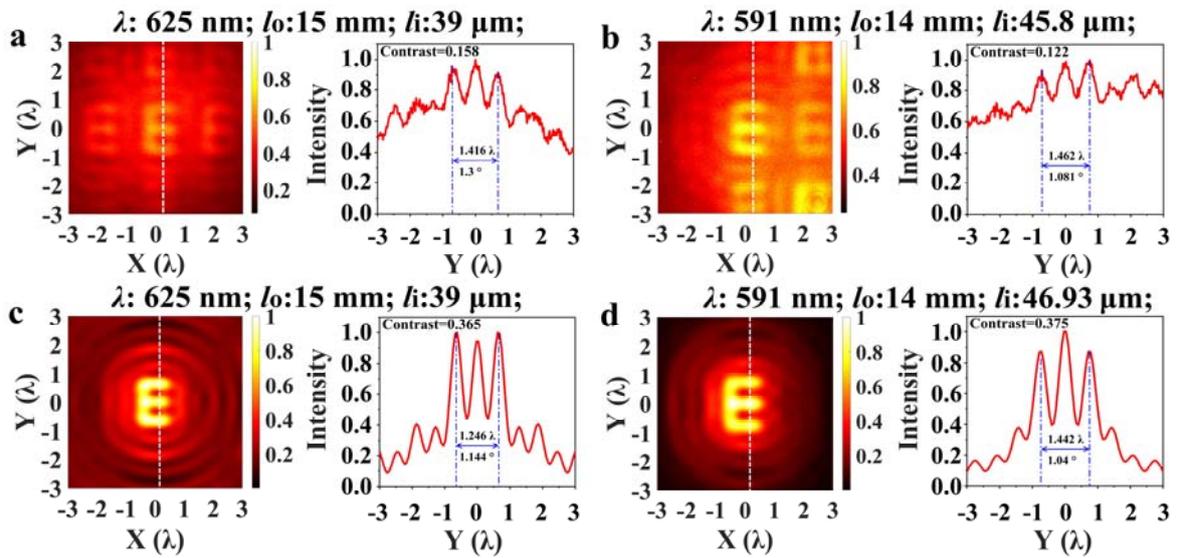

**Figure 9.** The imaging results of the proposed SLAM singlet at different illumination wavelengths of 625 nm and 591 nm. The experimental results at wavelengths of (a) 625 nm and (b) 591 nm; the numerical results at wavelengths of (c) 625 nm and (d) 591 nm.

To investigate the bandwidth of the fabricated SLAM singlet, the imaging experiments have also been conducted at two different wavelengths of 625 nm and 591 nm. **Figure 9a and 9b** present the experimentally obtained images at the two different wavelengths with the minimum resolvable angle of 0.306º and 0.327º for different objective distances of $l_o$ = 15 mm and $l_o$ = 14 mm respectively. The best image plane located at $l_i$ =39 μm and at $l_i$ = 45.8 μm. The image contrasts are 0.158 and 0.122 respectively. Therefore, they are greater than the value required by the Rayleigh criterion and the images are resolvable at both wavelengths under test. The corresponding angular magnifications are calculated to be 2.12 and 1.65 respectively. **Figures 9c and 9d** present the corresponding numerical results, which give better contrast values of 0.4 and 0.367. The corresponding angular magnifications are 2.02 and 1.614. Again, the experimental and numerical results demonstrate a good agreement with each other. It is worth emphasizing that, although the lens is optimized at wavelength of $\lambda$=632.8 nm, it both theoretically and experimentally demonstrates a super-resolution imaging performance at wavelengths of 625 nm and 591 nm, indicating a certain bandwidth of such a SLAM lens.

## ■ CONCLUSION

In conclusion, we have proposed a new way to realize super-resolution optical imaging by angular magnification. We have also proposed a new class of diffractive lenses with angular magnification greater than 1. Our concept has been theoretically and experimentally proved by specially designed diffractive lenses with angular magnification. The results demonstrate a more than two-fold enhancement in angular resolution, compared with the conventional angular resolution limit. Although, the demonstrated lenses have angular magnification of approximate 2, higher angular magnification is also possible. It provides a promising way to overcome the challenge in breaking through the conventional optical diffraction limit and solve the problem from the aspect of the optical devices and systems themselves. This idea of angular magnification and the new class of diffractive lenses with angular magnification should find their application in far-field super-resolution imaging, especially, for super-resolution imaging of objects in far distance, such as telescopes and optical remote sensing.

## ■ ASSOCIATED CONTENT

**Data Availability Statement**

The data that support the findings of this study are available from the corresponding authors upon reasonable request.

**Supporting Information.**

none

## ■ AUTHOR INFORMATION

**Corresponding Author**

Gang Chen - *Key Laboratory of Optoelectronic Technology and Systems (Chongqing University), Ministry of Education, School of Optoelectronic Engineering, Chongqing University,174 Shazheng Street, Shapingba, Chongqing 400044, China*
E-mail: gchen1@cqu.edu.cn

**Funding**

**Notes**

The authors declare no competing financial interest.

## ■ ACKNOWLEDGMENTS

Authors would like to extend sincere thanks to Prof. Ting Jiang and Prof. Yuanpeng Zou at School of Foreign Languages and Cultures, Chongqing University, for their assistance in language polishing.